\documentclass[aps, superscriptaddress, prd, 10pt, twocolumn, floatfix ]{revtex4}
\usepackage{amsmath, amssymb, amsfonts}
\usepackage{hyperref}
\usepackage{graphicx}
\usepackage{color}
\graphicspath{{./}{images/}}

\newcommand{\figref}[1]{Fig:~\ref{#1}}
\newcommand{\tabref}[1]{Table:~\ref{#1}}
\newcommand{\npart}[0]{$N_{\mathrm{part}}$}
\newcommand{\fm}[0]{\mathrm{fm}}

\begin{document}
\title{Mapping the Proton's Fluctuating Waistline}
\author{Christopher~E.~Coleman-Smith} \email{cec24@phy.duke.edu}
\affiliation{Department of Physics, Duke University, Durham, NC 27708-0305, USA}
\author{Berndt~M\"uller} \email{muller@phy.duke.edu}
\affiliation{Department of Physics, Duke University, Durham, NC 27708-0305, USA}
\affiliation{Brookhaven National Laboratory, Upton, NY 11973, USA}
\date{\today}

\begin{abstract}
  We discuss a mechanism for the apparently universal scaling in the high-multiplicity tail of charged particle
  distributions for high energy nuclear collisions. We argue that this scaling behavior originates from rare
  fluctuations of the nucleon density. We discuss a pair of simple models of proton shape fluctuations. A
  ``fat'' proton with a size of $3$~fm occurs with observable frequency. In light of this result, collective flow 
  behavior in the ensuing nuclear interaction seems feasible. We discuss the influence of these models on the
  large $x$ structure of the proton and the likely influences on the distribution of initial state spatial
  eccentricities $\epsilon_{n}$.
\end{abstract}

\maketitle

\section{Introduction}

We seek to understand high multiplicity events in p+Pb collisions at the LHC for which flow like properties
have recently been observed \cite{Chatrchyan:2013nka, Chatrchyan2013795}. 
The properties of events in the large \npart~tails of these multiplicity distributions resemble in many respects those of
Pb+Pb events at the same multiplicity. We propose that the tail of the p+Pb multiplicity distribution arises
from long-lived (on the collision time scale) quantum fluctuations in the colliding proton's wavefunction, as
opposed to fluctuations in the Pb nucleus or fluctuations in the final state particle production process.

Our argument is based on the hypothesis that the wave function of the nucleon  includes configurations that are
so spatially extended that their inelastic cross section is much larger than the average. These
fluctuations correspond to relatively low energy excitations of the proton in the comoving frame, which are
vastly time dilated in the reference frame of the Pb nucleus. As such they can be considered as approximately
frozen during the entire p+Pb collision, except for perturbations caused by the interactions with nucleons
in the Pb nucleus. 

Having a larger geometric size, it is natural to expect that the
incident proton will have a much larger cross section with the nucleus
when it finds itself in one of these configurations. As a result, more
energy will be deposited and more particles will be produced. Such
cross section fluctuations in hadron collisions have a relatively long
history of study \cite{Good:1960ba, Kopeliovich:1981pz,
  Blaettel:1993ah, Bialas:2006qf, Frankfurt:2008vi}. What is most
important for the interpretation of the observed collective flow-like
properties of the high multiplicity events, however, is that the
energy will be deposited over a much larger transverse area, which
makes the validity of a hydrodynamical description
\cite{Bozek:2011if,Bozek:2012gr,Bozek:2013df,Bozek:2013uha,Qin:2013bha}
of the following expansion more credible.

In the following, we will consider two alternative models for the spatial structure of the large-size
configurations of a highly boosted nucleon. The first model is based on the flux-tube model of
quark confinement (we call this the ``stringy'' nucleon). The second is a pion-cloud model, in which the
nucleon is surrounded by one or several soft virtual pions (we call this the ``cloudy'' nucleon). We will
argue on the basis of existing data for the antiquark distribution in the nucleon that the probability of
finding the nucleon surrounded by a cloud of four pions is of the order of $P(4\pi) \sim 10^{-6}$ and thus
should be abundantly sampled in the CMS experiment, which recorded an event sample corresponding to $6\times
10^{10}$ minimum bias events.

We start with a discussion of multiplicity fluctuations induced by fluctuations in the nucleon-nucleon
cross section, introduce two physical models for these fluctuations and finally develop models of the spatial
eccentricities arising from them.

\section{Multiplicity Fluctuations}

In the recent papers \cite{Alvioli:2013vk, Rybczynski:2013mla} the
authors consider fluctuations in the total nucleon-nucleon cross
section $\sigma_{\rm NN}$ arising from color fluctuations in the
initial nuclear densities along with the usual contributions from the
varying number of participating nucleons.

We reproduce some simple arguments which show that large geometric cross sections favor a large
number of nucleon-nucleon interactions. We shall set aside impact parameter fluctuations in \npart~ and only
consider contributions arising from a fluctuating nucleon cross section $\sigma$. Following the
optical Glauber model we consider the incident proton as a cookie-cutter punching out a tube of
cross sectional area $\sigma$ from the target nucleus. We define \npart~ as the number of
nucleons in this tube and take it to be Poisson distributed with mean 
\begin{equation}
\bar{n}(\sigma) = \sigma \rho L,
\end{equation}
where $\rho = 0.138~ \fm^{-3}$ is the nucleon density per unit volume and $L \approx 10$~fm 
is the length of the nucleus as seen by the incident proton in a central p+Pb collision. 
Then the probability of observing a given \npart~ is
\begin{equation}
p(N_{\mathrm{part}} ) = \frac{(\sigma \rho L)^{N_{\mathrm{part}}} }{N_{\mathrm{part}}!}\exp(-\sigma \rho L ). 
\end{equation}
Taking the average value of $\langle \sigma \rangle = \sigma_{\rm NN} = 4.803\; \fm^2$ \cite{Totem2013}
then $E[N_{\mathrm{part}}] = \bar{n}(\sigma_{\rm NN}) = 6.73$.
\begin{figure}[htb!]
  \centering
  \includegraphics[width=0.4\textwidth]{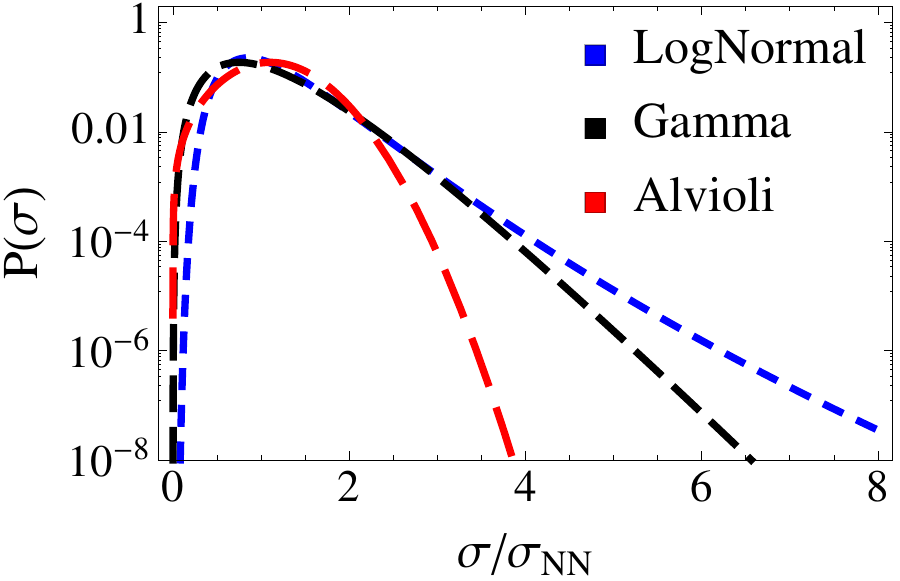}
  \caption{\label{fig:sig-dist-plot} (Color Online) Proposed probability distributions for fluctuations in the total cross section $\sigma_{\rm NN}$}
\end{figure}
Let us consider distributional forms for $\sigma$ along with that presented in \cite{Guzey:2005tk,
  Alvioli:2013vk}, we fix the mean of the proposed distributions to the average $\langle \sigma \rangle =
\sigma_{\rm NN} = 4.803\; \fm^2$. We pick two probability distributions to model the fluctuations of the cross
section: a gamma distribution and a log normal (see \figref{fig:sig-dist-plot}).  The densities are
\begin{align}
  p_{\mathrm{Alvioli}}(\sigma) &= \rho \frac{\sigma}{(\sigma + \sigma_0)} \exp(-\frac{\left(\sigma/\sigma_0  - 1\right)^2 }{\Omega^2}),\\
  &\rho= 0.363, \quad \Omega=0.69, \quad \sigma_0 = 4.80~\mathrm{fm}^2,\notag\\
  p_{\mathrm{gamma}}(\sigma) &= \frac{\sigma^{k-1} \exp \left(-\frac{\sigma}{\theta}\right)}{\theta^k \Gamma (k)}, \\
  &\theta = \frac{\langle \sigma_{\rm NN} \rangle}{k}, \quad k = 4.0, \notag\\
  p_{\mathrm{log normal}}(\sigma) &= \frac{1}{\sigma \delta \sqrt{2 \pi}} \exp \left( - \frac{(\log(\sigma) - \sigma_{NN})^2}{2\delta^2} \right),\\
  &\delta = 0.428, \notag
\end{align}
\begin{widetext}
We fix the values of $k$ and $\delta$ in the gamma and log-normal distributions so that both of the proposed
distributions have the same variance. The Miettenen-Pumplin relation \cite{Miettinen:1978jb} connects the scaled variance of
$P(\sigma)$ to the ratio of single inelastic and elastic cross-sections at $t=0$
\begin{equation}
  \label{eqn-mpump}
  \int P(\sigma) \left( \frac{\sigma}{\sigma_{NN}} - 1 \right) ^2\; d \sigma \equiv \omega_{\sigma} = \left. \frac{ d\sigma( p + p \to X + p )/dt }{ d\sigma( p + p \to P + p )/dt } \right |_{t=0}
\end{equation}
\end{widetext}
Our proposed distributions have $\omega_{\sigma} = 0.25$ which is consistent with current experimental results. 

\begin{figure*}[ht!]
  \centering
  \includegraphics[width=0.3\textwidth]{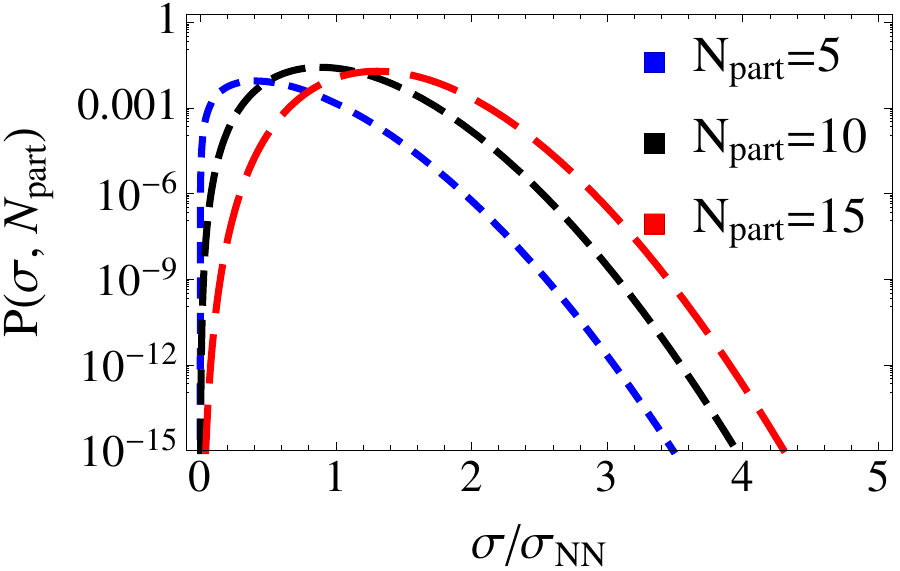}
  \includegraphics[width=0.3\textwidth]{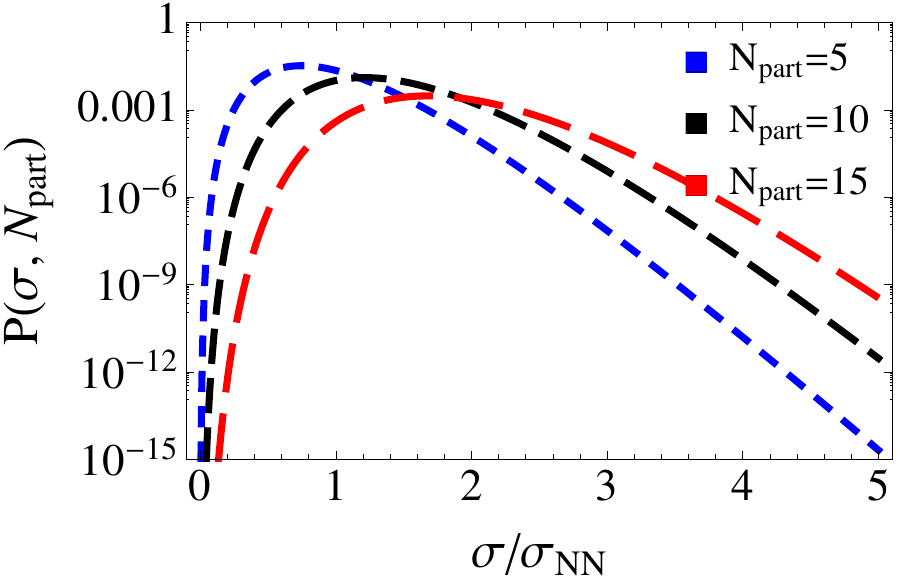}
  \includegraphics[width=0.3\textwidth]{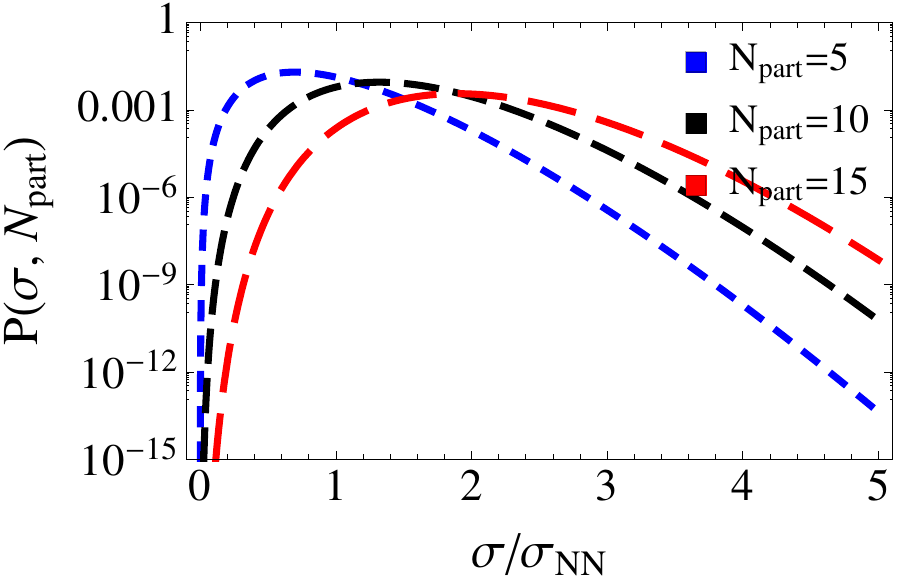}
  \caption{\label{fig:joint-plots} Joint probability distributions for $\sigma$ and \npart~ at values of fixed
    \npart. From left to right the Alvioli, gamma and Gaussian distributions are shown. These results do not include the effects of impact parameter fluctuations or nucleon-nucleon correlations. }
\end{figure*}

\begin{figure}[ht!]
  \centering
  \includegraphics[width=0.3\textwidth]{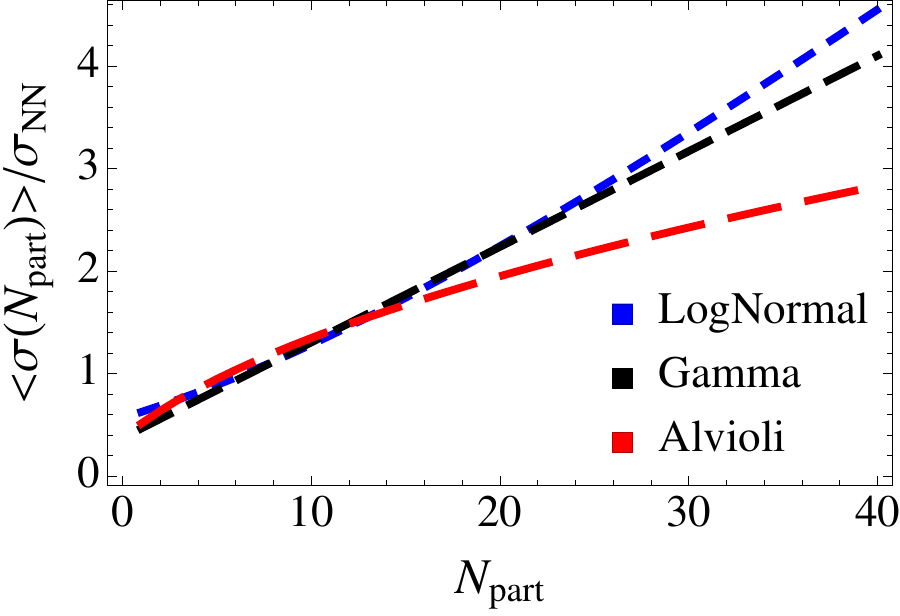}
  \caption{\label{fig:sigma-averaged} The average cross section as a function of the number of participants
    for each of the proposed cross section distributions. These results do not include the effects of impact
    parameter fluctuations or nucleon-nucleon correlations. }
\end{figure}

\begin{figure}[ht!]
  \centering
  \includegraphics[width=0.3\textwidth]{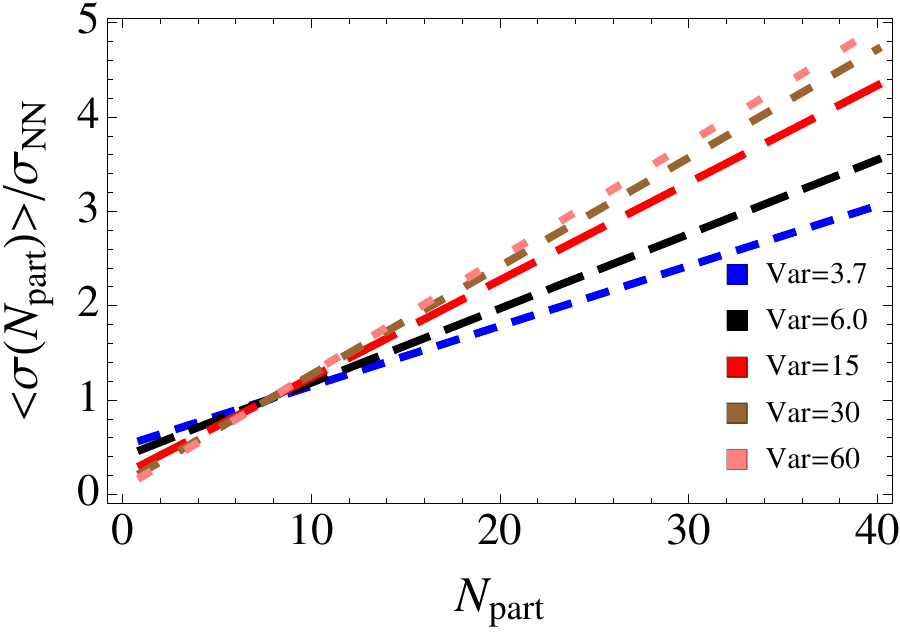}
  \caption{\label{fig:sigma-averaged-gamvar} The average cross section as a function of the number of
    participants for the gamma cross section distribution with constant mean and increasing variance.}
\end{figure}

The joint probability distributions of \npart~ and $\sigma$ are shown for some fixed values of \npart~ in
\figref{fig:joint-plots}. From these figures it is clear that large fluctuations in the cross section $\sigma$
are more likely to contribute at larger values of \npart.  We compute the average cross section
$\hat{\sigma}(N_{\mathrm{part}})$ for each of the proposed distributions
\begin{equation}
\hat{\sigma}(N_{\mathrm{part}}) = \frac{\int_{0}^{\infty} \sigma\, p(\sigma, N_{\mathrm{part}})\; d\sigma }{\int_{0}^{\infty} p(\sigma, N_{\mathrm{part}})\; d\sigma}.
\end{equation}
These effective cross sections are shown in \figref{fig:sigma-averaged} as a function of the
number of participants. The effective cross section grows roughly linearly with the number of
participants. Events with large \npart~ are more likely to be events with a large cross section and thus a
large effective proton area. We show the influence of the variance of the proposed cross section distributions
in \figref{fig:sigma-averaged-gamvar}, a larger variance enhances the effective cross section for a given
number of participants.

Having established that fluctuations in the cross section can be selected by requiring large fluctuations 
in the number of participants, let us now consider some simple models of these fluctuations.

\section{The Stringy Model}

In the stringy model we model the fat proton as three valence quarks connected by color flux tubes. This
phenomenological model is inspired by results from quenched lattice QCD which show that at even relatively 
modest valence quark separations the gluon field in a nucleon localizes into flux tubes \cite{Bissey:2009gw,
Bissey:2006bz}. In 3-body problems it is often convenient to use Jacobi coordinates
\begin{align}
  u &= x_2 - x_1, \quad p_{u} = \frac{1}{2} (p_1 - p_2),\notag\\
  v &= (x_2 + x_1)/2 - x_3, \quad p_{v} = \frac{1}{3} (p_1 + p_2 - 2 p_3), \notag\\
  w &= (x_1 + x_2 + x_3)/3, \quad p_{w} = p_1 + p_2 + p_3.
\end{align}
In the center-of-mass (CM) frame $p_{w} = w = 0 $. Neglecting spin effects, we can write a
wave equation for this system as 
\begin{equation}
\left[p_1^2 + p_2^2 +p_3^2 + V(x_1, x_2, x_3)^2 \right] \Psi = E^2 \Psi,
\end{equation}
where $V(x_1,x_2,x_3)$ is the inter-quark potential \cite{Sakurai:1967}. Here 
we approximate this potential as a linear confinement potential with string constant $k$ in the limit of
very spatially extended configurations. We assume a star-like configuration of flux tubes converging on
the CM of the quark configuration:
\begin{equation} 
  V(x_1, x_2, x_3) = k \left( \left| \frac{u}{2}+\frac{v}{6} \right| +  \left| \frac{u}{2}-\frac{v}{6} \right| + \left| \frac{2v}{3} \right| \right).
\end{equation}
Neglecting cross terms in the large spatial extent regime we approximate the potential for convenience as
\begin{equation} 
V(x_1, x_2, x_3)^2 = k^2(u^2+v^2) .
\end{equation}
The wave equation then takes the form
\begin{equation}
\left[ \left( 2 p_{u}^2 + \frac{3}{2}p_{v}^2\right) + k^2( u^2 + v^2) \right] \Psi = E^2 \Psi.
\end{equation}
The solution for the wave function is
\begin{equation}
\Psi(u,v) = N \exp \left( - \frac{ku^2}{2\sqrt{2}} - \frac{kv^2}{\sqrt{6}} \right).
\end{equation}
From the normalization requirement
\begin{equation}
1 = \int_{0}^{\infty}  \Psi^2\; u^2 v^2 du dv,
\end{equation}
we obtain $N^2 = \frac{16}{\pi 3^{3/4}} k^3 $. The mass of the nucleon in this simple model is 
\begin{equation}
E^2 = \left( \sqrt{2} + \sqrt{3/2} \right) k = 0.53\;\mbox{GeV}^2.
\end{equation}

\begin{figure}[htb!]
  \includegraphics[width=0.3\textwidth]{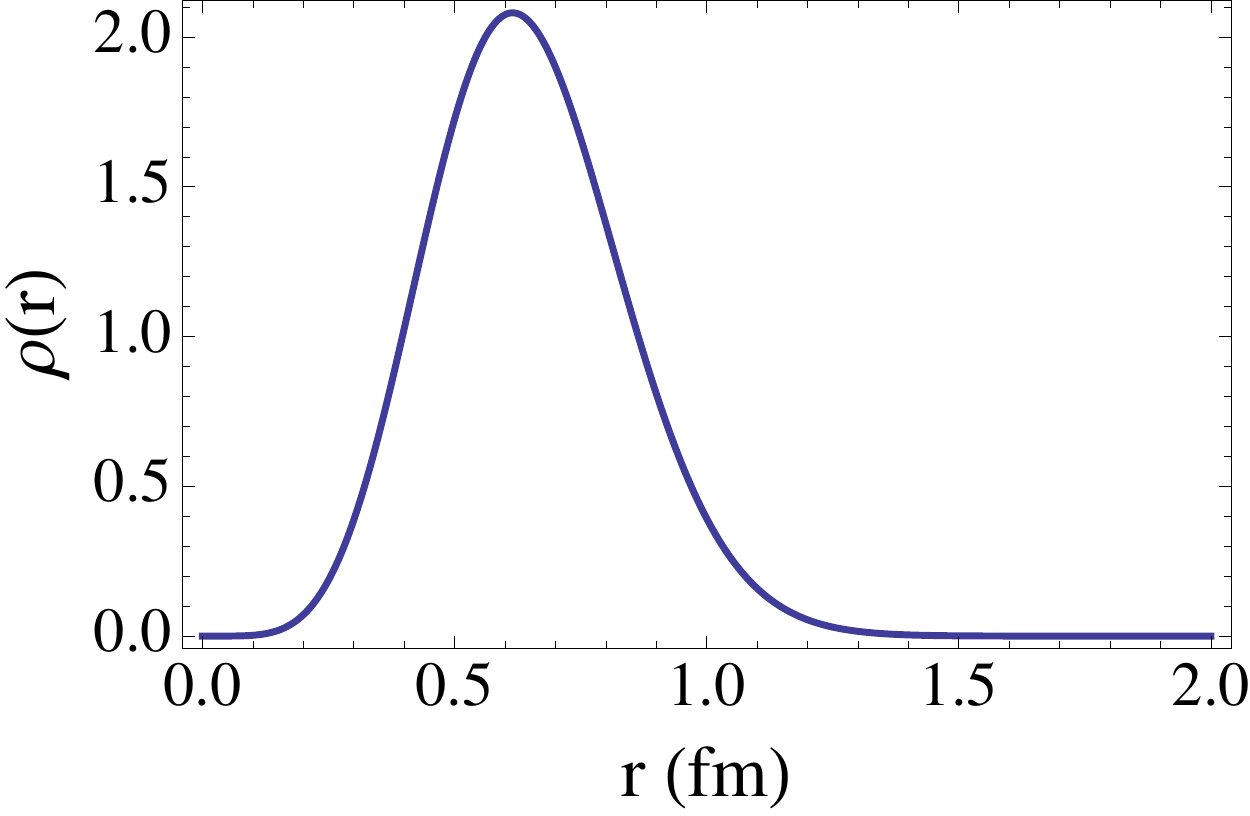}
  \caption{\label{fig:radius-dist} (Color Online) The probability distribution for the mean square radius of the extended nucleon $\rho(r)$}.
\end{figure}

The mean square radius of the system is
\begin{align}
  \langle r^2 \rangle &= \int u^2 v^2 \Psi^2 \frac{1}{6}(3 u^2 + 4v^2) \; du dv, \notag\\
  &= \frac{3^{1/4}\sqrt{6+\frac{7\sqrt{3}}{2}}}{2k} = \frac{2.28541}{k} ,
\end{align}
taking a string constant $k = 1~\mathrm{GeV}/\fm$ we obtain the root-mean-square (RMS) radius of these
configurations $\sqrt{\langle r^2 \rangle} = 0.674~\fm$.  The fraction of configurations of radius $\rho(r)$ can be computed from 
\begin{align}
  \rho(r^2) &= \int \Psi^2 \delta\left(r^2 - \left(\frac{u^2}{2} + \frac{2 v^2}{3}\right) \right) u^2 v^2 \; du dv,
\end{align}
this is plotted in \figref{fig:radius-dist}.

\begin{figure}[htb!]
  \centering
  \includegraphics[width=0.3\textwidth]{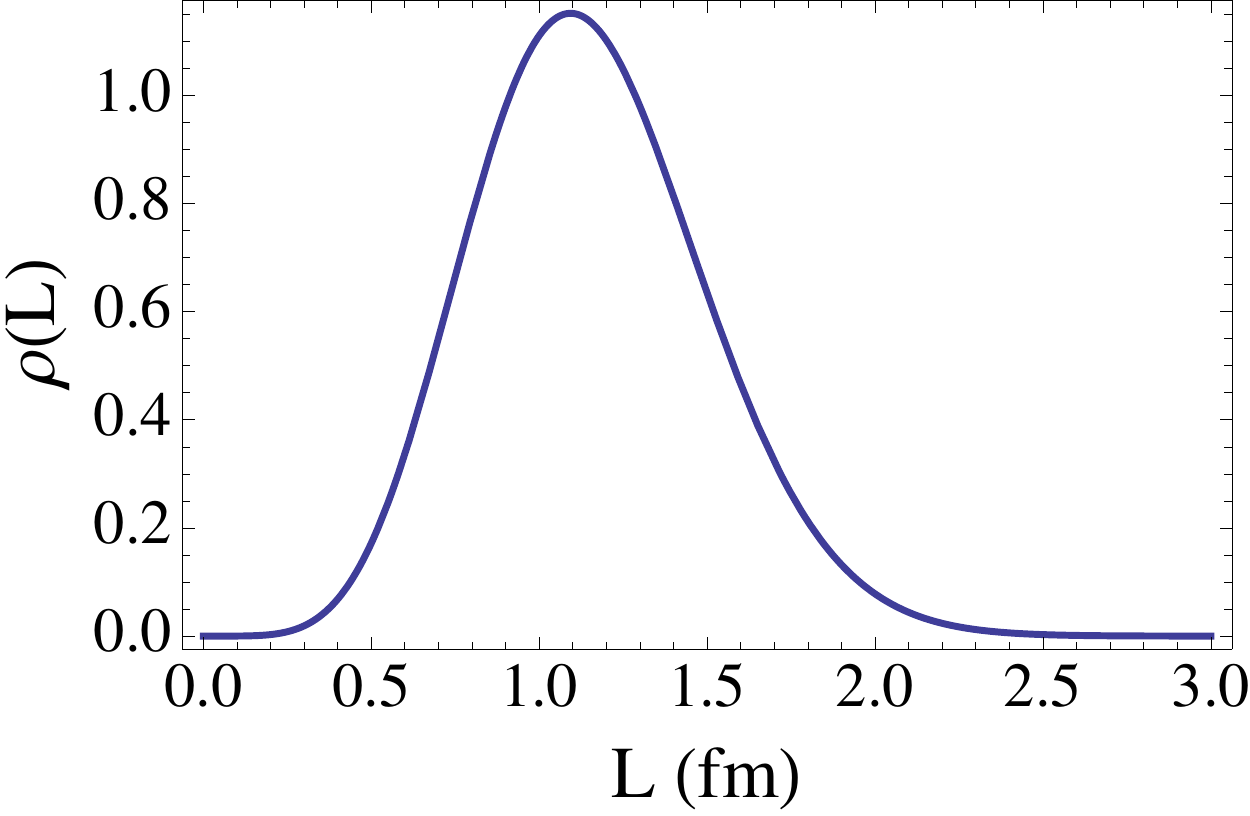}
  \caption{\label{fig:rho-r-plot} The probability distribution for the total flux tube length in the limit
    of a very extended proton}
  \includegraphics[width=0.3\textwidth]{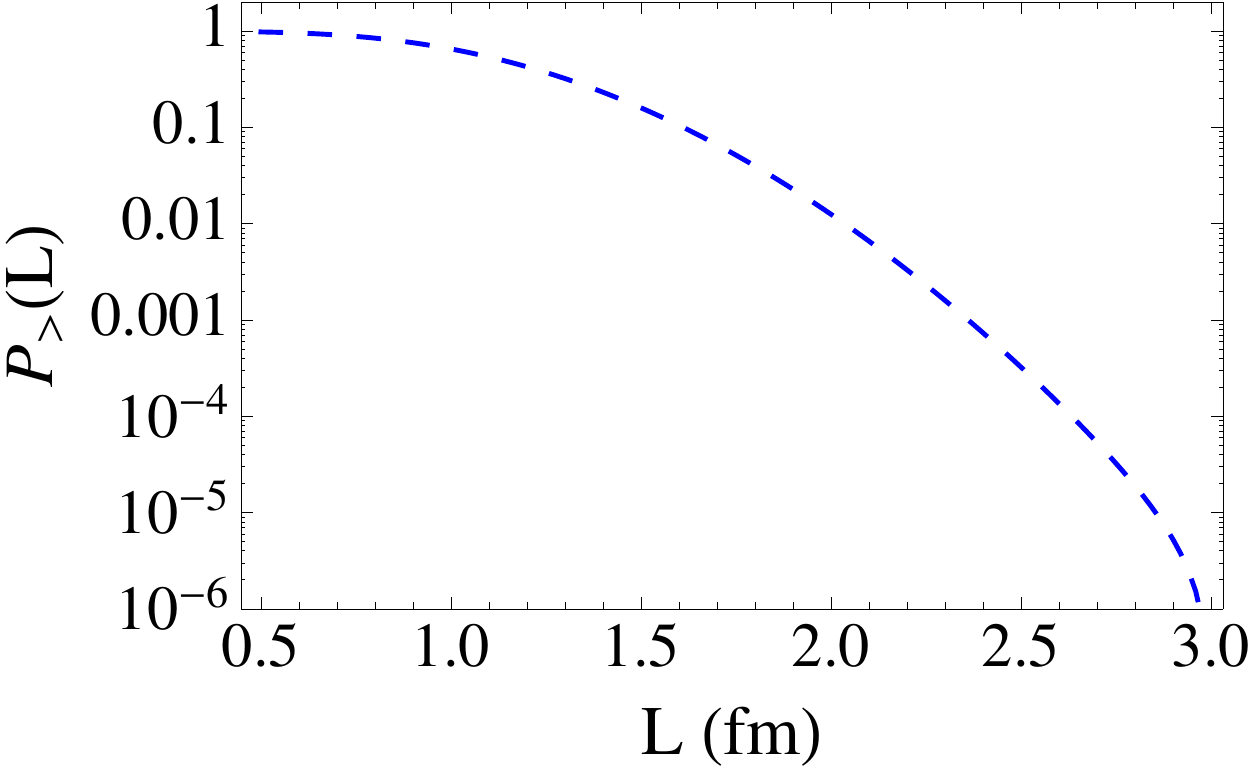}
  \caption{\label{fig:pgreater-plot} The probability for the total flux tube length to be greater than $L$ in the limit
    of a very extended proton}
\end{figure}

The fraction $\rho$ of configurations with total flux-tube length $L = u + v$,
\begin{equation}
  \rho(L) = \int  \Psi^2 \delta(u + v - L)\; u^2 v^2 du dv ,
\end{equation}
is shown in \figref{fig:rho-r-plot}.  The average total flux tube length is $\langle L \rangle = 1.155~\fm$. The
probability of the total flux tube length exceeding a certain value of $L$, 
\begin{equation}
P_>(L) = \int_L^\infty \rho(L') dL' ,
\end{equation}
is shown in \figref{fig:pgreater-plot}. Configurations with very long flux tubes occur with observable
frequency, for instance, we would expect there to be approximately $10^4$ events with with 
$L > 2.8~\fm$ in the CMS p+Pb data.

\section{The Cloudy Model}

There is a non vanishing probability for a proton to produce a virtual pion via the transition $p \to n \pi^+$ 
or $p \to p \pi^0$. Isospin symmetry dictates that $P(p \to n \pi^+) = 2 P(p \to p \pi^0)$. The proton can also 
produce a virtual pion and simultaneously excite itself into one of the states of the $\Delta$-resonance: 
$p \to \Delta\pi$. As this transition requires an additional 300 MeV of energy, we neglect this contribution
here, but it would need to be taken into account in a more complete treatment.

Since the configuration
with a single virtual pion contains either a neutron or a proton, it can spawn another virtual pion by the
same mechanism. Assuming that the consecutive pion production processes are independent then the 
number of virtual pions $N_{\pi}$ accompanying the proton is given by a Poisson distribution with mean 
given by the average number of virtual pions $\langle n_{\pi} \rangle$. The probability of finding the 
incident nucleon accompanied by a cloud of $N_{\pi}$ pions is thus:
\begin{equation}
  \label{eqn:poisson-npi}
P(N_{\pi}) = \frac{\langle n_{\pi} \rangle ^{N_{\pi}} \exp(-\langle n_{\pi} \rangle)} {N_\pi !}.
\end{equation}
Experimental information about the virtual pion cloud of the nucleon is obtained, e.~g.\, from exclusive 
pion production in electron scattering off the nucleon, or the measurement of the isovector component 
of the antiquark distribution in the nucleon \cite{Horn:2006tm, DeTroconiz:2001wt, Amendolia:1984nz}. Here we focus on the second method.

\subsection{The $\bar{d}/\bar{u}$ asymmetry}

Parton distribution functions $f_i(x,Q^2)$ (PDFs) \cite{Brock:1993sz} give the unnormalized probability of 
finding a parton of species $i$ with a given momentum fraction $x$ in a proton at a given scale $Q$. 
The distributions are normalized so that
\begin{align}
  \int_{0}^{1} f_{u} (x) - f_{\bar{u}}(x)\; dx &= 2,\\
  \int_{0}^{1} f_{d} (x) - f_{\bar{d}}(x)\; dx &= 1,\\
  \int_{0}^{1} x f_q(x) + x f_{\bar{q}}(x) + x f_{g}(x)\; dx &= 1. 
\end{align}
The first two integrals fix the number of valence quarks of each flavor in the proton, the third sum 
ensures conservation of the total momentum of the proton.

The Gottfried sum rule is given in terms of the second nucleon structure function $F_2^{N}(x,Q^2) = \sum_{a} x
f_{a} (x, Q^2)$,
\begin{align}
  \mathfrak{S}_{g} &= \int_0^{1} (F_2^{p} - F_{2}^{n}) \frac{dx}{x},\notag \\
  &= \frac{1}{3} + \frac{2}{3} \int_0^{1} \left[ f_{\bar{u}}(x) - f_{\bar{d}}(x) \right]\; dx .
\end{align}
The naive value for this sum would be $\mathfrak{S}_g = \frac{1}{3}$, based on the notion that
sea quarks arise from the splitting of gluons, implying that the antiquark distribution functions in the 
nucleon are flavor symmetric. However several experiments have found a non vanishing net flavor 
asymmetry in the distribution of sea quarks \cite{Towell:2001nh}. A review of the theory of this 
asymmetry can be found in \cite{Kumano:1997cy}. We shall pursue the idea that the asymmetry is 
the consequence of the presence of a cloud of virtual pions as developed in
\cite{PhysRevD.53.2586, PhysRevD.44.717}. The E866 results \cite{Towell:2001nh} give 
$\int_{0}^{1} \left[ f_{\bar{d}}(x) - f_{\bar{u}}(x) \right]\; dx = 0.118 \pm 0.012$. 
If we interpret the asymmetry as arising solely from the production of virtual pions we can set 
$P(p \to n \pi^+) = 0.118$. Considering isospin symmetry, this leads to the conjecture 
$\langle n_{\pi} \rangle = \frac{3}{2} \times 0.118 = 0.177$.

\subsection{Antiquark distribution in the nucleon}

\begin{figure}[ht!]
  \centering
  \includegraphics[width=0.4\textwidth]{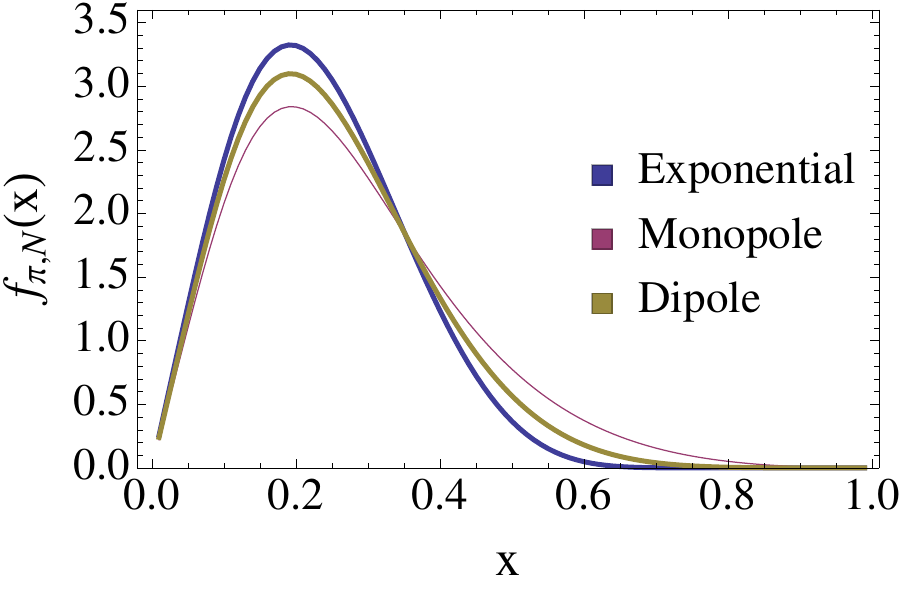}
  \caption{\label{fig:pion-dist} The virtual pion momentum distribution function $f_{\pi,N}$ computed for each of the form factors.}
\end{figure}

Following \cite{PhysRevD.53.2586, PhysRevD.44.717} we can write down the contribution to the overall proton
light antiquark PDF of a single virtual pion. This is given in terms of the convolution of the light-cone
momentum distribution of a virtual pion $f_{\pi,N}$, the probability of finding a pion with momentum fraction
$y$, and the pion antiquark PDF $g_{\bar{q}}(x/y, Q)$:
\begin{equation}
x f_{\bar{q}}^{(1)}(x,Q) = \mathcal{C}^2  \int_x^{1} dy f_{\pi,N}(y) \frac{x}{y} g_{\bar{q}}\left(\frac{x}{y}, Q\right). 
\end{equation}
Where $\mathcal{C}$ is the associated isospin Clebsch-Gordan coefficient,
\begin{equation}
f_{\pi, N}(y) = -\frac{g_{\pi NN}^2}{16 \pi^2} y \int_{-\infty}^{tm} \frac{ -t}{\left(t - m_{\pi}^2\right)^2} | F_{\pi NN}(t)|^2,
\end{equation}
$F_{\pi NN}(t)$ is the nucleon-nucleon-pion form factor and $t_{m} (y) = -M_{N}^2 \frac{y^2}{(1-y)}$ is the
maximum invariant momentum transferred to the pion. In the literature the following form factors are suggested
\cite{Kumano:1997cy, PhysRevD.43.3067}
\begin{align}
  F_{\pi NN}^{\mathrm{monopole}} &= \frac{\Lambda_m^2 - M_{\pi}^2}{\Lambda_{m}^2 -t}, \notag\\
  F_{\pi NN}^{\mathrm{dipole}} &= \left(\frac{\Lambda_d^2 - M_{\pi}^2}{\Lambda_{m}^2 -t}\right)^2, \notag\\
  F_{\pi NN}^{\mathrm{exp}} &= \exp\left(\frac{t-M_{\pi}^2}{\Lambda_{e}^2}\right).
\end{align}
Setting $\Lambda_d = 0.8$, $\Lambda_m = 0.62 \Lambda_d$ and $\Lambda_e
= 1.28 \Lambda_m$ as suggested by Kumano \cite{Kumano:1997cy} we
obtain the pion distribution shown in \figref{fig:pion-dist}. Here we
have chosen $g_{\pi NN}$ such that the distribution $f_{\pi, N}(y)$ is
normalized to one. This allows us to interpret $f_{\pi, N}(y)$ as the
probability for finding a pion at a given momentum fraction $y$
\emph{given that there is a pion present in the nucleon} as opposed to
setting the value from experimental data and interpreting it as the
unconditional probability of finding a pion with momentum fraction $x$ in the nucleon.  
As can be seen, the choice of form factor does not have a significant influence
on the pion momentum distribution. From here on we use the dipole form
as it is the median curve in \figref{fig:pion-dist}. 
The average pion momentum is relatively independent of the form factor
\begin{equation}
  \langle x_{\pi} \rangle = \frac{\int_0^{1} x f_{\pi,N}(x)\; dx } {\int_0^{1} f_{\pi,N}(x)\; dx } = 0.234.
\end{equation}

Let us consider the probability of observing a light antiquark conditioned on the number of pions present in the
system. The conditional probability of observing a light antiquark given that there are no pions, $P_{\bar{q}}(x,Q | N_{\pi} = 0)$, is
\begin{align}
   x P_{\bar{q}}(x ,Q| N_{\pi} = 0)  &=  x f_{\bar{q}}(x,Q) P(N_{\pi} = 0),
  \end{align}
where for simplicity we are taking the nucleon PDF $f_{\bar{q}}(x,Q)$ as being defined in the absence of virtual pions. 
The conditional probability for observing a light antiquark with momentum fraction $x$ given that there is a
single pion accompanying the proton is
\begin{align}
  & x P_{\bar{q}}(x,Q | N_\pi = 1) = \int_x^{1} dy\, f_{\pi,N}(y) \left\{ \frac{x}{y} g_{\bar{q}}\left(\frac{x}{y}, Q\right) \right. \notag \\
  & \left. \qquad + \frac{x}{1-y} f_{\bar{q}}\left(\frac{x}{1-y}, Q\right) \right\} P(N_{\pi} = 1).
\end{align}
The probability for finding a light antiquark with momentum fraction $x$ and there being a single pion in the
system is the sum of terms representing the probability of finding the light antiquark \emph{within the pion} and
the probability of finding the light antiquark \emph{in the proton} given that the pion has taken away a fraction
$y$ of the proton's total momentum. Similarly we can write down the conditional probabilities for configurations
with more virtual pions. As an example, we give the result for $N_\pi = 2$:
\begin{widetext}
\begin{align}
  x P_{\bar{q}}(x,Q | N_\pi = 2) &= \int_{x}^{1}\; dy_{1} \int_{x}^{1-y_{1}}\; dy_{2}  f_{\pi,N}(y_1) f_{\pi, N}\left(\frac{y_2}{1-y_1}\right) \notag\\
  &  \times \left\{ \frac{x}{y_1} g_{\bar{q}}\left(\frac{x}{y_1}, Q\right) + \frac{x}{y_2} g_{\bar{q}}\left(\frac{x}{y_2}, Q\right) 
      + \frac{x}{1-y_1 - y_2} f_{\bar{q}}\left(\frac{x}{1-y_1 - y_2}, Q\right)\right\} P(N_{\pi} = 2).
\end{align}
\end{widetext}

\begin{figure}[htb]
  \centering
  \includegraphics[width=0.4\textwidth]{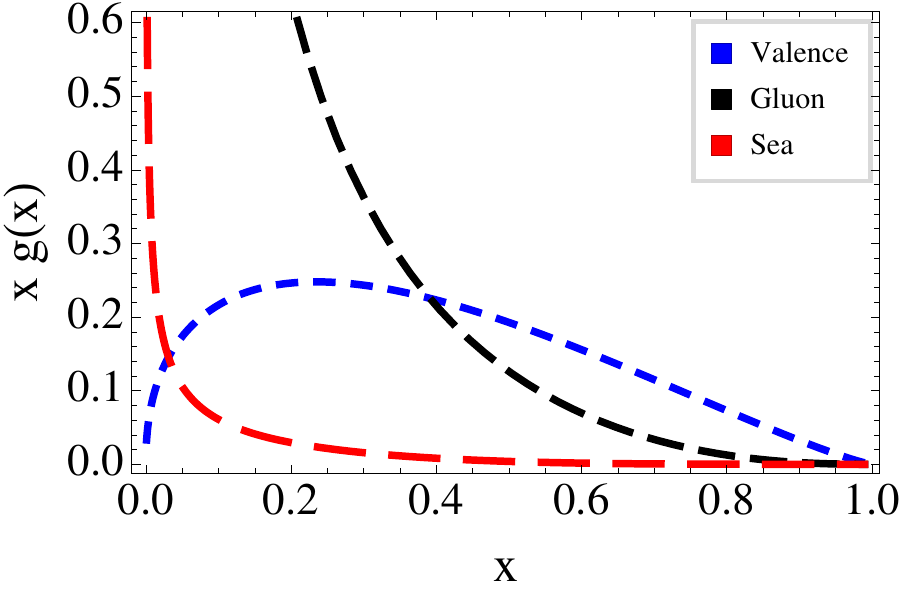}
  \caption{\label{fig:pipdf} The Gl\"uck {\em et al.} \cite{Gluck:1991ey} HO
    pion PDFs evaluated at $Q=10$~GeV, for valence and sea quarks and
    gluons. The average momentum fractions for each species are respectively $0.155,0.023,0.511$.}
\end{figure}

In our evaluation of these expressions we use the parameterization given by Gl\"uck {\em et al.} \cite{Gluck:1991ey} 
for the pion PDF $g_{\bar{q}}$. For these PDFs, at $Q=10$~GeV the average valence antiquark momentum 
fraction is $\langle x_{\bar{q}} \rangle_{\pi} = \int_{0}^{1} x g_{\bar{q}}(x,Q)\; dx = 0.155$. 
For reference we plot the valence, sea and gluon PDFs of the pion in \figref{fig:pipdf}.

\begin{figure}[htb]
  \centering
  \includegraphics[width=0.4\textwidth]{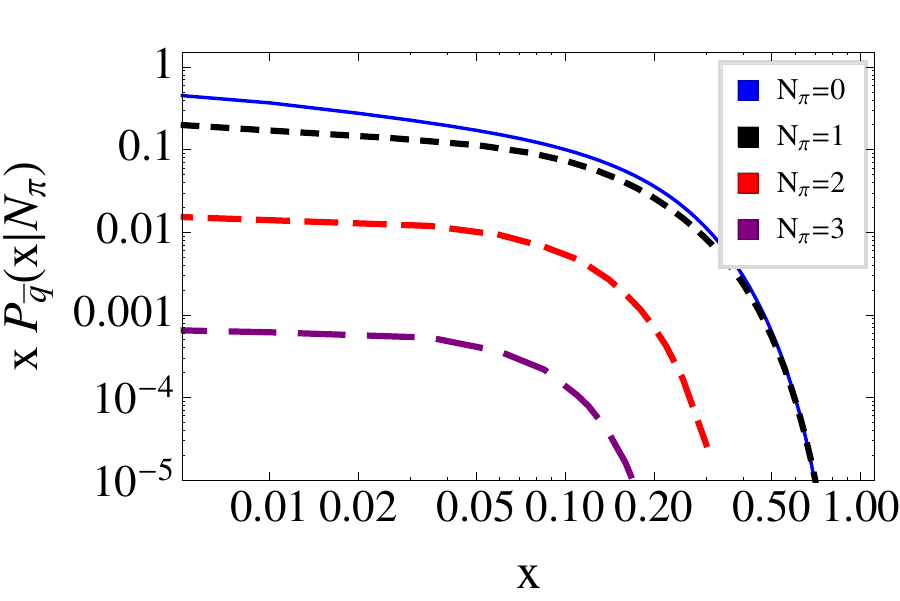}
  \caption{\label{fig:modpdf-contrib} The contribution to the light antiquark parton distribution functions
    from configurations with a given definite number of pions. The number of pions ranges from 0 to 3, computed at $Q = 10$~GeV.}
\end{figure}

\begin{table}
  \centering
  \begin{tabular}{|l|r|c|}
    \hline
    $N_{\pi}$ & $P(N_{\pi})$ & $\int dx\, P_{\bar{q}}(x,Q | N_{\pi})$ \\
    \hline
    0 & 0.889~~~ & 2.292\\
    1 & 0.104~~~ & 0.747\\
    2 & 0.00618 & 0.068 \\
    3 & 0.00024 & 0.0027\\
    4 & $7.17 \times 10^{-6}$ & \\
    \hline
  \end{tabular}
  \caption{\label{tab:modpdf} The probability of finding $n$ pions along with the integrated light quark PDF,
    computed at $Q = 10$~GeV. The integral over the PDF is cut off at $x_{min} = 0.001$.}
\end{table}

\begin{figure}[htp]
  \centering
  \includegraphics[width=0.4\textwidth]{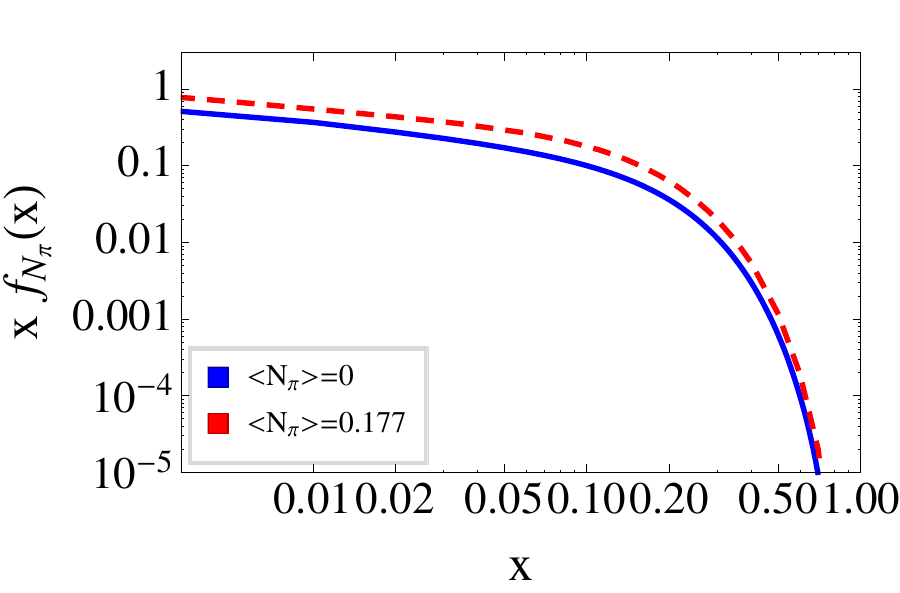}
  \includegraphics[width=0.4\textwidth]{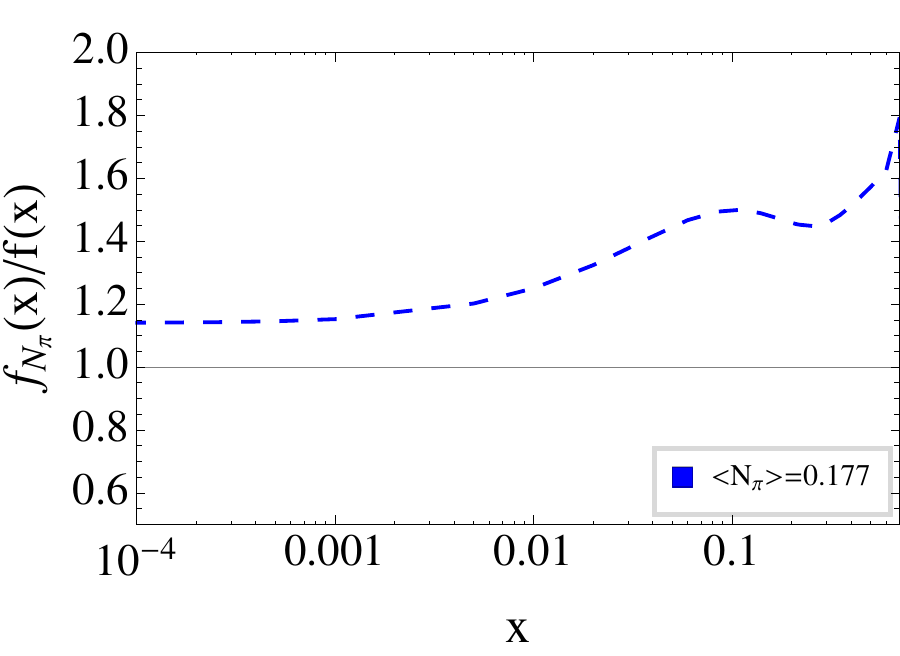}
  \caption{\label{fig:modpdf} The modified light antiquark PDF (top)
    plotted with the unmodified PDF. The ratio of the modified light
    antiquark PDF to the unmodified case (bottom), computed at $Q = 10$~GeV}
\end{figure}

We tabulate the probabilities of finding $n$ pions in the physical proton along with the integral of the modified 
PDF in \tabref{tab:modpdf}.
The contributions from configurations with different numbers of virtual pions to the antiquark distribution 
are shown in \figref{fig:modpdf-contrib}. The contributions die off quickly with $N_{\pi}$, the higher order 
terms contribute to successively smaller ranges in $x$ due to the conservation of the total momentum of 
the proton. The modified PDF including effects from up to three pions,
\begin{equation}
\label{eqn:full-mod}
x \tilde{f}_{\bar{q}}(x, Q) = \sum_{n=0}^{3} x P_{\bar{q}}(x, Q | N_{\pi} = n) ,
\end{equation}
is shown in \figref{fig:modpdf}. 

\section{Phenomenology}

\subsection{Effects on Hard Scattering}

An observable consequence of the stringy proton model would be an enhancement of gluon jet production over
quark jets in high multiplicity p+Pb events. We expect that the gluon density in a ``fat'' nucleon will be
enhanced at moderate to large $x$, as almost all of the energy in the proton now resides in the gauge field
contained in the flux-tubes. This implies that the momentum fraction carried by the valence quarks must be
shifted to smaller values of $x$. We expect that the localization of the valence quarks and the enhanced
large-$x$ gluon distribution will have a non trivial feed-down to saturation physics at small $x$. However a
more detailed calculation is needed to address the small-$x$ physics associated with a high multiplicity p+Pb
event. We expect that the total cross-section fluctuations arising from this model would scale like
fluctuations in the total area of the nucleon, this is set by $r^2= \frac{1}{6}\left(3 u^2 + 4v^2\right)$
\begin{equation}
p\left(\frac{\sigma}{\sigma_{NN}}\right) \propto p\left(\frac{r^2}{\langle r^2 \rangle}\right) 
\end{equation}
where $p(r_{\mbox{rms}})$ is plotted in \figref{fig:radius-dist}.

In the case of the cloudy proton model, the presence of virtual pions in the ``fat'' proton serves to enhance 
the antiquark PDF at large values of $x$. This enhancement must be accompanied by a shift of the light 
quark distribution to smaller values of $x$. This could lead to an observable enhancement of hard quark-antiquark
annihilation, expressed as enhanced Drell-Yan pair-production or as enhanced W-boson production 
in high multiplicity events relative to a minimum bias baseline. 

We note that in both models the valence quark distribution will be shifted to lower values of $x$, implying
reduced production of very hard jets initiated by valence quark scattering. As a consequence, in the cloudy 
nucleon model the gluon sea, and as a secondary effect the isoscalar quark sea will be much enhanced, 
while in the stringy nucleon model the isovector quark sea will be enhanced with little or no increase in
the gluon sea. This difference should, in principle, serve as an observable distinction between the two 
models. Of course, in reality, both mechanisms may contribute to the ``fat'' proton configurations.

To estimate the significance of these modifications we consider the cloudy nucleon model.  We can 
compare the average momentum fraction carried by an antiquark in the modified and unmodified situations,
\begin{equation}
  \label{eqn:meanx}
\langle x_{\bar{q}} \rangle = \frac{\int_{x_{\mathrm{min}}}^{1} x f_{\bar{q}}(x)\; dx}{\int_{x_{\mathrm{min}}}^{1} f_{\bar{q}}(x)\; dx},
\end{equation}
we use a lower cut-off of $x_{\mathrm{min}} = 0.001$.
In terms of the antiquark distribution inside the virtual pion we can estimate 
\begin{equation}
  \langle x_{\bar{q}}^{\pi} \rangle \simeq \langle x_{\pi} \rangle \langle x_{\bar{q}} \rangle_{\pi} = 0.234 \times 0.168 = 0.0393,
\end{equation}
using data from \figref{fig:pion-dist} and \figref{fig:pipdf}. Directly integrating the MSTW PDFs for nucleon sea antiquarks we find
\begin{equation}
  \langle x_{\bar{q}}^{N} \rangle = 0.0119.
\end{equation}
This means that the antiquarks contributed by virtual pions carry, on average, three times the longitudinal
momentum than the antiquarks contained in the parton sea of an average proton. This difference should 
be possible to observe if the population of protons with a virtual pion can be significantly enhanced by
selecting high-multiplicity p+Pb events.
The fully modified PDF \eqref{eqn:full-mod} including the effects of up to three virtual pions gives
\begin{equation}
  \langle x_{\bar{q}}^{N,\mathrm{Mod}} \rangle = 0.0173,
\end{equation}
a value $3/2$ times larger than the unmodified case. The virtual pions make a significant contribution to the nucleon PDF. 
We can expect some modification to hard processes as a result.
  
We refrain here from making quantitative predictions for hard scattering phenomena accompanying
high multiplicity p+Pb events, because these will certainly depend sensitively on the possible trigger 
conditions, which are not known to us.  We also are concerned that the sophistication of the models 
of the ``fat'' proton explored here, especially the ``stringy'' proton model, is insufficient to make reliable
quantitative predictions for the effective parton distributions associated with a given multiplicity window.

\subsection{Eccentricity Distributions}

How else can we physically distinguish between these two toy models? By considering their influence on the
parton distributions we have examined the fat proton in a longitudinal section. We now attempt to build models
of the transverse structure of the portly proton. We numerically sample the spatial eccentricity coefficients
$\epsilon_2, \epsilon_3$ from density distributions generated in the spirit of each of the models. If the energy
deposited in a proton-nucleus collision thermalizes and the tiny fireball expands hydrodynamically, these
spatial eccentricities may reasonably be expected to be reflected in the Fourier coefficients $v_{n}$ of the 
final-state flow . 

We compute the eccentricities for an event with a transverse density profile $\rho$ as
\begin{equation}
  \epsilon_{n} = \frac{\int \rho(r,\phi) r^2 \cos(n \phi - n \Phi_n) r dr d\phi}{\int \rho(r,\phi) r^3 dr d\phi},
\end{equation}
where the event plane angle $\Phi_{n}$ for the $n-$th moment is
\begin{equation}
  \Phi_{n} = \frac{1}{n} \arctan \left( \frac{ \int \rho(r,\phi) r^2 \sin(n\phi) r dr d\phi}{ \int \rho(r,\phi) r^2 \cos(n\phi) r dr d\phi} \right).
\end{equation}

We generate events for the pion cloud model with $N$ pions as follows, where $N$ is drawn from the Poisson
distribution \eqref{eqn:poisson-npi} . For each event we sample the radial locations of the $N$ pions about
the proton from an exponential distribution. The pion angular positions are sampled uniformly. The exponential
radial distribution is motivated by the Yukawa model, we consider several values of the rate constant
$\lambda$ for this distribution. In \cite{PhysRevD.53.2586} the authors carry out a more advanced calculation
along the same lines as our cloudy model, including the effects of the $\Delta \pi$ channel. The dominant
contribution of the pion cloud to the antiquark distribution arises from pions with an average momentum of
$\langle P_{\pi} \rangle \simeq 0.8$~GeV, although this calculation is carried out at a slightly lower
virtuality scale $Q^2 = 1\;\rm{GeV}^2$ this result provides a reasonable estimate for the mean radial position
of pions around the proton. We set our the average pion radial position to be $\lambda = \frac{1}{\langle
  P_{\pi} \rangle}$.

A Gaussian kernel with width $\sigma_{\pi} = 1/\sqrt{6}$~fm is convolved against the resulting points. This
kernel width is chosen so that $r_{\pi} = \sqrt{\frac{2}{3}} r_{p}$. We take the radius of the proton as
defining the $2\sigma$ distance from the center, i.e the probability of finding any density outside of this
radius is $< 5\%$. Finally a density representing the proton is placed at the origin with a smearing kernel
width of $\sigma_{p} = 0.5$~fm.

\begin{figure*}[h]
  \centering
  \includegraphics[width=0.8\textwidth]{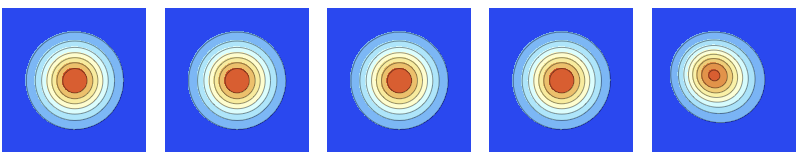} \\
  \includegraphics[width=0.8\textwidth]{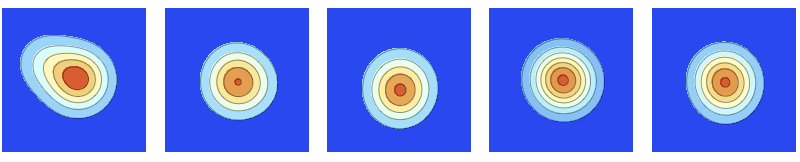}   
  \caption{\label{fig:pion-dists} Contour distributions of the proton and pion-cloud density (arb) in the
    transverse plane, the width of each plot is $3$~fm. Each plot is a single event sampled from the
    ensemble. The top row shows events with $\langle N_{\pi} \rangle = 0.1778$ the calculated value, the
    bottom row shows events with $\langle N_{\pi} \rangle = 4$.}
\end{figure*}

Density plots of a few typical events from the pion model are shown in \figref{fig:pion-dists},
here the plots have a width of $3/2$~fm. The central proton tends to dominate the density but the effects of the
outlying pions are visible. We consider one ensemble with the average number of pions set to the physical value of $\langle N_{\pi} \rangle = 0.1778$  and one with $\langle N_{\pi} \rangle = 4$ to illustrate the effects of large fluctuations.

For the stringy model we sample the absolute values of the Jacobi parameters $u,v$ normally with some width
$\sigma_{\rm string}$ such that the average total flux tube length is $\langle \rho \rangle = 1.009$~ fm,
to match the values we computed above. The angles made by $u,v$ in the transverse plane are sampled uniformly,
the positions of the three quarks can then be reconstructed.
\begin{figure*}[h]
  \centering
  \includegraphics[width=0.9\textwidth]{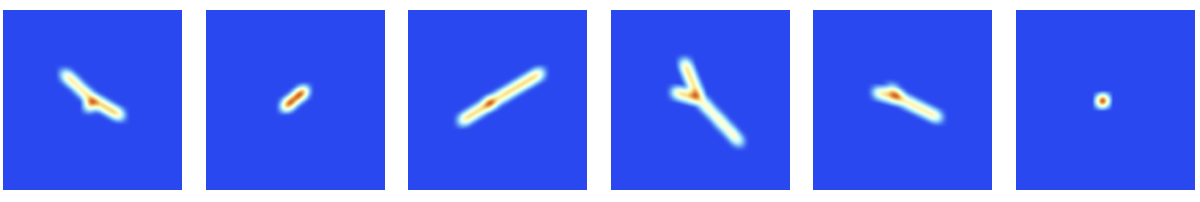}\\
  \includegraphics[width=0.9\textwidth]{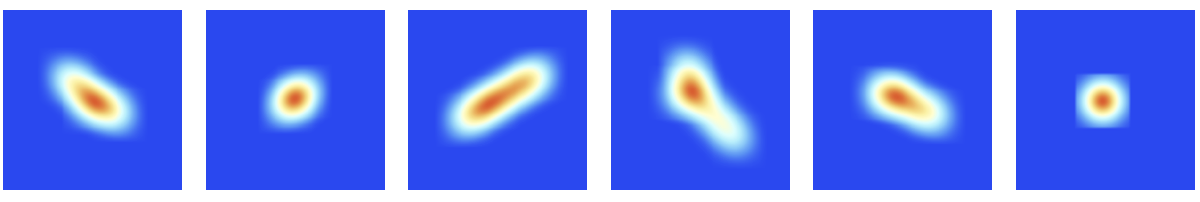}
  \caption{\label{fig:string-dists} Distributions of the stringy density (arb) in the transverse plane, the width of each plot is $2$~fm. The top row shows strings with a width $0.1$~fm the bottom row shows strings with a width $0.3$~fm.}
\end{figure*}
The flux tube density profile is generated by convolving the resulting line segments with a Gaussian profile. We
consider two ensembles, a ``thin'' and ``fat'' set of events with widths $w_{\mathrm{string}} = 0.1, 0.3$~fm
respectively. Some typical events are shown in \figref{fig:string-dists}. These plots are $3$~fm in width,
long two-legged configurations tend to dominate.

\begin{figure*}[ht!]
  \centering
  \includegraphics[width=0.3\textwidth]{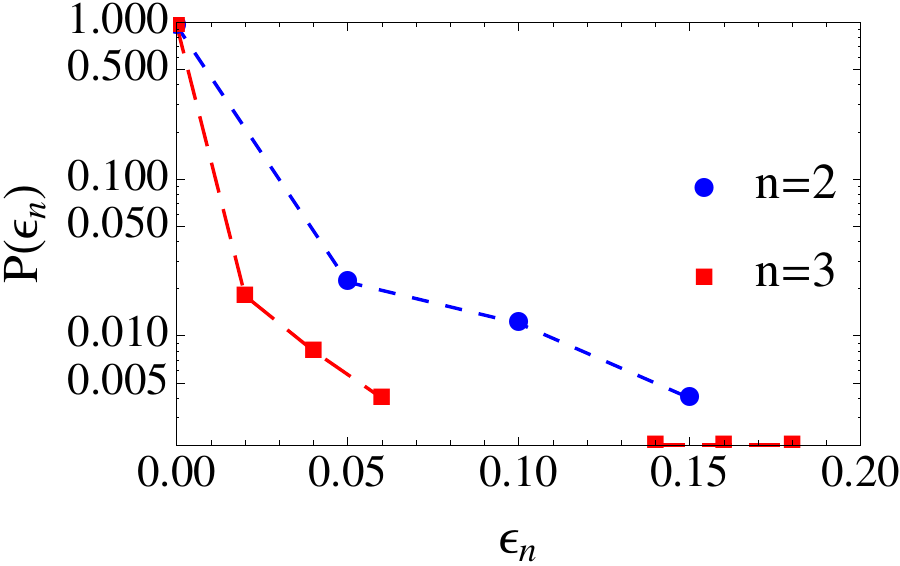}
  \includegraphics[width=0.3\textwidth]{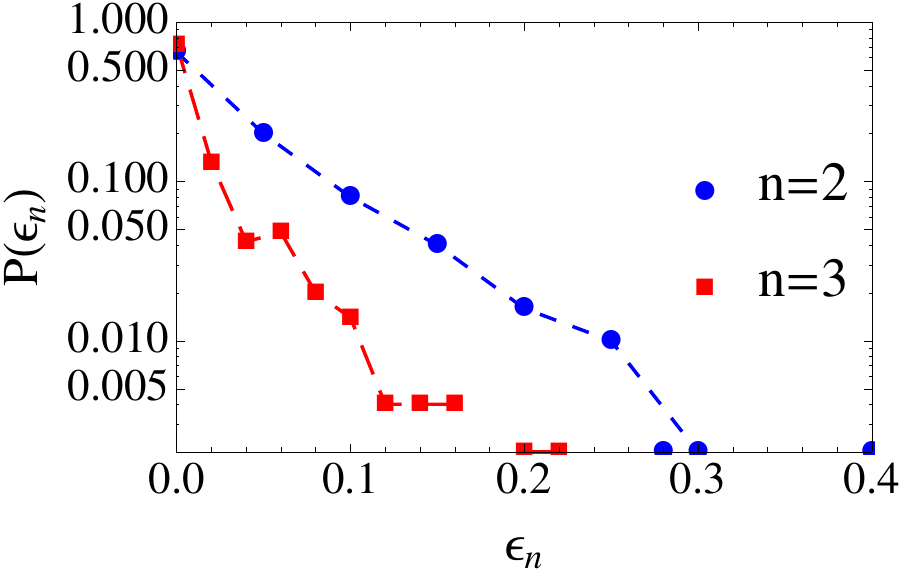}
  \caption{\label{fig:eps-dists-pion} Distributions of $\epsilon_2, \epsilon_3$ for $500$ events generated
    from the pion-cloud. The left figure shows the results for the physical value $\langle N_{\pi} \rangle =
    0.1778$, the right figure shows the results for $\langle N_{\pi} \rangle = 4$. Note that impact parameter
    fluctuations are not included. }
\end{figure*}

\begin{figure*}[hbt!]
  \centering
  \includegraphics[width=0.3\textwidth]{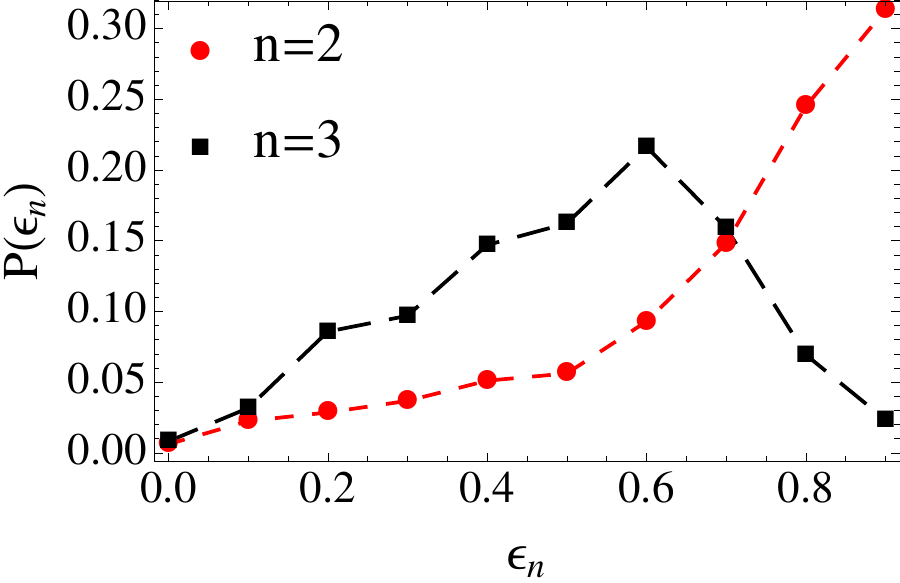}
  \includegraphics[width=0.3\textwidth]{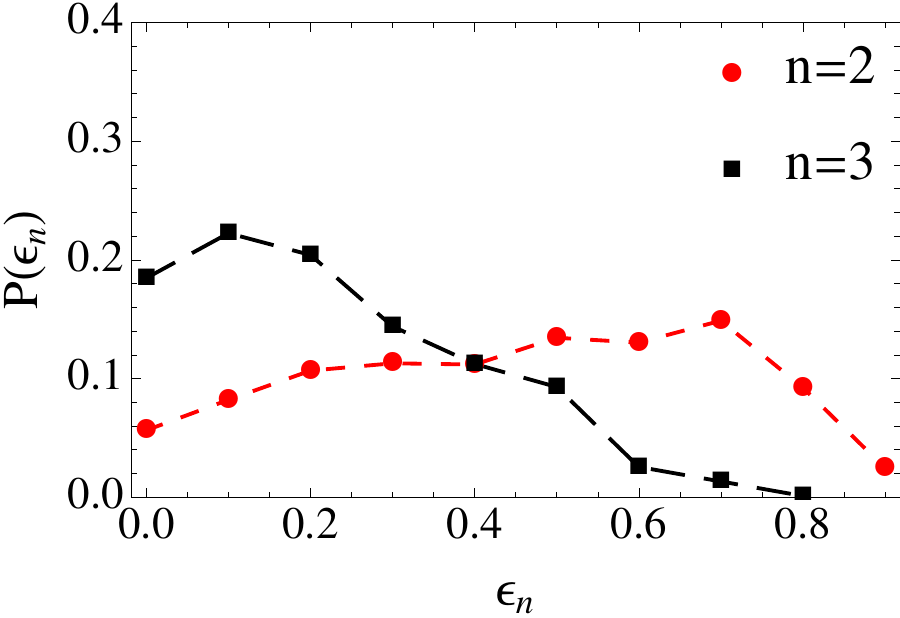}
  \caption{\label{fig:eps-dists-string} Distributions of $\epsilon_2, \epsilon_3$ for $1000$ events generated
    from the stringy models. The string width is $0.1$~fm in the left figure (thin) and $0.3$~fm in the right
    figure (fat). Note that impact parameter fluctuations are not included.}
\end{figure*}

Histograms of the $\epsilon_2/\epsilon_3$ distributions for the pion cloud and stringy models are shown in
\figref{fig:eps-dists-pion} and \figref{fig:eps-dists-string}. The pion model with a realistic average number
of pions per nulceon gives an appreciably non-zero eccentricity distribution, this is strongly enhanced in for
the large pion cloud case.

Either choice of flux tube width leads to strong enhancements in the $\epsilon_2$
spectrum at large eccentricities and to a nontrivial $\epsilon_3$ spectrum, the wider string model shows less
dramatic results as the smearing reduces the geometric influence of the string profile.

\section{Summary and Outlook}

Fluctuations in the nucleon-nucleon cross section can induce large fluctuations in the number of participants
in a central p+Pb event. The apparent universality of the large \npart\ tails of p+Pb, Pb+Pb and p+p collisions
suggests these fluctuations arise from a spatially over-extended, or ``fat'', proton wave function. A natural
consequence of this extended proton size and its concomitant large cross section is an enhanced collision 
volume in such p+Pb events. The larger volume reduces spatial density gradients and thus makes a
hydrodynamical description of the evolution of the reaction more likely to be valid. 

We have proposed two phenomenological models for the large-size configurations of the proton, one based 
on color flux tubes and one on virtual pion production. Each model leads to modified large $x$ physics in the 
initial state of the p+Pb collision relative to minimum bias events. 
Qualitatively, the stringy proton model predicts enhancement of the gluon PDF, 
while the cloudy proton model predicts an enhancement of the light antiquark PDFs. 
It would be interesting to view these models as different initial seeds for small $x$ saturation
physics. The stringy model's extended ``valence'' gluon configuration  is likely to give rise to a
substantially different color glass than that arising from the pion cloud which effectively has many 
more valence (anti-)quarks.

In proton-nucleus collisions the conjectured ``fat'' proton configurations have obvious consequences 
for the transverse energy density distribution in the initial state and its Fourier moments $\epsilon_{n}$. 
The much enhanced initial transverse extent of the fireball makes the application of hydrodynamical
models for its expansion more credible, because it implies a larger Knudsen number. Since the
distribution of eccentricities is significantly different for the two models considered here,
measurements of final-state ``flow'' coefficients $v_{n}$ for high multiplicity p+Pb events
will shed some light upon which of these models is more realistic.

\begin{acknowledgments} 
  We acknowledge support by DOE grant DE-FG02-05ER41367.  CCS would like to thank D.~Velicanu and
  I.~C.~Kozyrkov for many helpful discussions.
\end{acknowledgments}

\bibliographystyle{h-physrev5} 
\bibliography{./fat-proton-refs}

\end{document}